# The Internet and Community Networks: Case Studies of Five U.S. Cities

John B. Horrigan, Senior Research Specialist

Pew Internet and American Life Project
1100 Connecticut Ave., NW
Suite 710
Washington, DC 20036

[www.pewinternet.org](www.pewinternet.org)
202.296.0019

# I. Introduction

The Internet's role in communities has long been a topic that brings out a spirit of optimism among city officials, community activists, and citizens. The Internet presents the chance to engage people in civic life in new ways, and may even bring previously apathetic or ignored populations into the active citizenry. The Internet also holds out the possibility of new economic opportunities for people and more effective service delivery for local governments. Time will tell whether this optimism is warranted, but citizens and public officials in cities throughout the country are taking steps today to realize future benefits associated with the Internet. In this paper, I examine how community institutions in five U.S. cities are responding to the various possibilities presented by the Internet, whether in the realm of access for low-income people, job training, or improved service delivery for government and non-profits. The cities are: Austin, Texas; Cleveland, Ohio; Nashville, Tennessee; Portland, Oregon and; Washington, D.C. The cities were chosen for the geographical diversity and variability in economic profiles and Internet penetration rates. An appendix presents economic data on the cities. For the case studies, city officials, community activists, and employees of community technology centers and other community-based organizations were conducted.[1]

The aim here will be two fold. The first is classification. By looking across a range of projects in different cities, one clear finding is that there is considerable variability across community technology projects. I will place the projects studies into several categories. Second, I will ask whether community technology projects have affected social capital in the cities studied. If social capital, as has been written, is all about social networks, and if community technology projects are intended to use information networks to carry out social goals, there is *a priori* reason to suspect that such projects affect social capital. I will first develop a framework for studying the Internet's impact on social capital, and then present findings from the five cities.

## II. SOCIAL CAPITAL AND THE INTERNET

### a. Definitions

Political scientist Robert Putnam characterizes social capital as "those features of social organization, such as trust, norms, and networks, that can improve the efficiency of society by facilitating coordinated actions."[2] In other words, social capital is all about social networks. Social capital can be thought of in two ways: *bridging* social capital and *bonding* social capital. Bridging social capital allows disparate groups in society to come together in ways they normally do not. The civil rights movement, which brought young Northern whites into contact with Southern blacks, is often cited as an example of bridging social capital. Bonding social capital refers to organizations that deepen ties

---

[1] See John B. Horrigan, forthcoming, "Cities, Cyberspace, and Social Capital: How Five U.S. Cities are Adapting to the Internet", Pew Internet & American Life Project, www.pewinternet.org, for further detail, including a listing of people interviewed.
[2] Robert Putnam, *Making Democracy Work: Civic Traditions in Modern Italy*. Princeton, NJ: Princeton University Press, 1993, p. 167.

among groups with a lot in common; country clubs are good examples of bonding social capital.

Sociologist Ray Oldenburg offers a somewhat different perspective on what generates social capital, saying that "third places" (i.e., neither home nor work, but places such as coffee shops or bars) are important glue for communities. These places are welcoming neutral ground for people, where conversation is the primary activity and status is unimportant. Such places offer people a comfortable environment in which to expand possibilities in their lives.[3] Although Oldenburg does not talk about social networks, "third places" certainly expand them for people who congregate there. These places facilitate both bridging and bonding social capital, as residents of a neighborhood can find out what is going on while also welcoming new people to the neighborhood or discovering what is going on in other parts of the city.

The specific way in which institutions facilitate cooperation—and thus build social capital—is through their affect on the cost of transactions.[4] If a group wants to organize the neighborhood to change the mind of City Hall, it is costly to marshal interest, settle on a message, and deliver it to elected officials. An institution such as a neighborhood association can reduce the costs of organizing. In both of these examples, the institutions amount to the "rules of the game" for carrying out transactions. In other words, the institutions are key sources of people and information for telling actors how things get done in a given environment and what the norms are for social cooperation.[5]

The Internet can play a role in reducing transactions costs in two ways. First, to the extent that a portion of these costs is informational, the Internet can play a role in reducing them. Whether through email or the Web, the Internet provides lots of information quickly and cheaply—information that could aid cooperation. Second, the Internet, due to its relative novelty in organizations, can serve as a catalyst to overcoming the friction that is part of any collective undertaking. This catalytic effect usually arises as organizations try to figure out how best to integrate the Internet into their missions. If the catalytic effect takes hold, it may result in the development of Internet content that furthers the missions of organizations. The net impact of the catalytic and "content" effects is to change the "rules of the game" that define how an organization functions.

### b. A Framework for Analysis

How can you tell when "rules of the game" are changing in an institution and, more importantly, if you can, how do you attribute it to the Internet? The answers have to do with foot traffic and content. With respect to foot traffic, the Internet may serve as an attractor for organizations, meaning that the presence of Internet connections may bring new people to a place who might not otherwise go there. This can inject new life into an organization by stimulating social networks. In this way, foot traffic is an indicator of the catalytic effect of the Internet on social capital formation.  It is the presence of the Internet that shapes social capital, as people establish new networks of contacts as they congregate at places where the Internet is.

---

[3] Ray Oldenburg, *The Great Good Place: Cafes, Coffee Shops, Bookstores, Bars, Hair Salons, and Other Hangouts at the Heart of a Community.* New York: Marlowe and Company, 3rd Edition, 1999.
[4] Putnam, *Making Democracy Work,* p. 179.
[5] Douglass North, *Institutions, Institutional Change, and Economic Performance.* New York: Cambridge University Press, 1990, p. 93.

Turning to content, Internet-driven projects may result in the creation of new Internet content that is devoted to addressing economic or community needs.  Such projects take the interplay between the Internet and social capital one step beyond foot traffic.  Rather than the Internet shaping social capital, as is the case when the Internet spurs new social networks, the presence of social capital is shaping the Internet through the creation of specialized content.  This is a stronger indicator of the connection between the Internet and social capital, because content creation only comes about if levels of trust about the Internet's potential have been established in the "foot traffic" phase of the Internet's development within an organization.

The "Internet as catalyst" theme, whereby the Internet's presence alters foot traffic, figures prominently in the five cities.  If a community organization decides to provide Internet access and training, the organization may draw new people to it.  This changes the character of the organization, while providing a different kind of place where people can gather—perhaps a new "third place" for some people.  The upshot may be more social capital, but due only in part to the Internet as a "network of networks."  The planning activities that lay the groundwork for exploiting the Internet in organizations result in higher levels of trust in the community than would otherwise be the case.

Content creation comes into play less frequently in this paper, but its impact is important when it is present.  When affordable housing providers come together in a city to develop a Web-based system to track the supply and condition of housing, this Internet content greatly improves operating efficiencies for clients.  When neighborhood non-profits help residents create Web pages for their home businesses, this reflects a growing level of trust in the neighborhood, and the content on the Web pages represents economic opportunities that benefit individuals and communities.  It takes time for content to translate into higher levels of trust in a community, but Internet-driven social capital is not likely to arise unless the initial catalytic effect from Internet planning translates into content.

The following figure lays out schematically the dynamic between the Internet and social capital, separating out the catalytic and content effects, but also showing how the two reinforce each other.  An existing level of social capital will influence the development of Internet content relevant to a particular community, but the presence of the Internet may spur planning activities whose stimulus to social networks builds social capital in the community.  This impact on social capital may in turn result in additional Internet content.  In the projects studied in the five cities, some fall clearly into the "catalytic" category, others in the "content " category, and some in between the two.

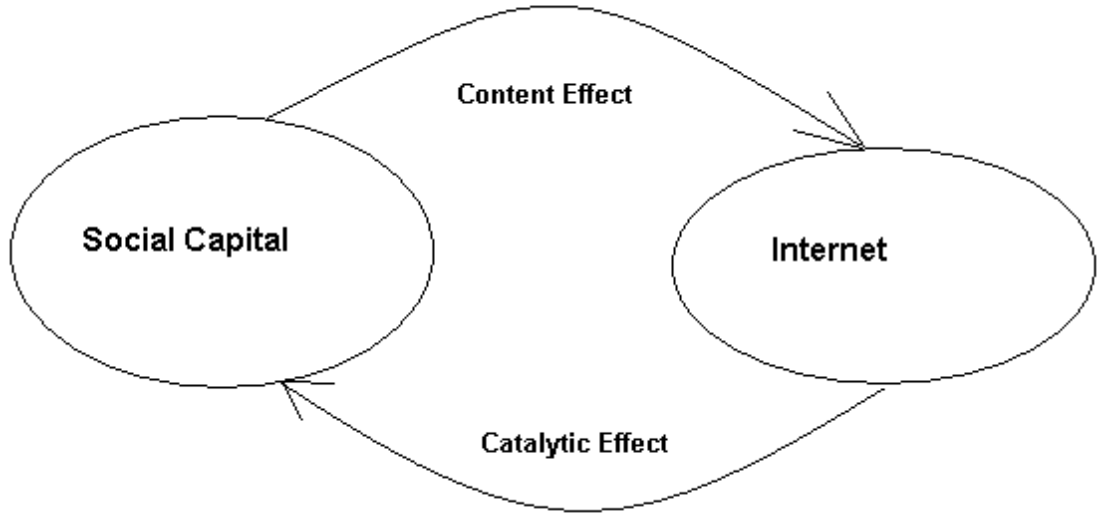

Others have pointed out the limitations in linking the Internet with social capital. As Putnam has written "[v]ery few things can yet be said with any confidence about the connection between social capital and the Internet."[6] Putnam acknowledges the potential for the Internet to build social capital, because the Internet is, after all, a network that connects people. He rightly concludes that "the Internet will not *automatically* offset the decline of more conventional forms of social capital, but it has that potential."[7] It is this potential—the ways in which communities are beginning to try to convert this potential into reality—which this paper charts.

---

[6] Robert Putnam, *Bowling Alone: The Collapse and Revival of American Community.* New York: Simon & Schuster, 2000, p. 170.
[7] *Ibid.,* p. 180.

# III. Findings from Five Cities

Community goals come in many shapes and sizes. Some places may suffer from a shortage of trained workers; others may suffer—at least in the minds of some community members—from a dearth of democratic participation in segments of the population. In some cities, committed individuals may shape the kinds of community initiatives that are undertaken. Given the variability in communities and their priorities, it is no surprise that there are different kinds of community Internet projects. What follows is a categorization of the community Internet projects in the five cities studied.

### a. Integrating access with social service delivery

Several cities have undertaken initiatives designed to use the Internet to improve the delivery of one of the most important of all social services—affordable housing. Matching the Internet with affordable housing is a natural for several reasons. Coordinating the delivery of affordable housing services—from construction to maintenance to leasing—is a complex undertaking that could benefit from the coordination that the Internet can provide. Finding ways to provide home Internet access to residents of affordable housing is also potentially very powerful. This opens up to these residents a host of options, such as access to job information, job training, or the ability to start a home-based business.

In Portland, Oregon, nearly 80% of all phone calls to social services agencies are placed to the wrong agency. That is, when a person calls Portland's housing bureau, four out of five times he will be referred to another agency for the service that he really needs; this is the case on average across all social service agencies in Portland. Through an ambitious program begun by the Bureau of Housing and Community Development, Portland plans to adopt a "travel agent" model for service delivery in the metropolitan area. With the support of a $480,000 grant from the U.S. Commerce Department's Technology Opportunities Program (TOP), the BHCD is creating a Web-based system to improve the delivery of housing and other services to Portland citizens. The goal is allow a person inquiring about a social service to be directly referred to the right provider, with service provision beginning immediately. For instance, someone seeking legal services who also needs and qualifies for affordable housing could immediately be told what is available, what must be done to apply, and the documentation needed to demonstrate eligibility. With home Internet access for residents of affordable housing, residents could not only find out about jobs or job training, but also enjoy the Internet's other benefits, such as email or online games.

The story is similar in Cleveland, where a group of non-profits organizations have developed a Web portal called T2K (on the Web at www.t2k.org). The scope of this program is ambitious in a way different from Portland. In addition to linking providers of affordable housing, T2K also links community development corporations with funders and in general improves the technological capabilities of neighborhood non-profits. T2K was developed with the help of a $525,000 grant from the federal TOP program.

### b. Education

Perhaps the most prominent rationale for encouraging Internet adoption among low-income populations is education. Without the Internet, the rarely disputed argument goes, those students lacking access will be adrift in our technologically sophisticated society. To some extent, education is a common theme in nearly every community access project, but several have it as a centerpiece.

In Austin, the Austin Learning Academy (ALA) is an after school learning program that has transformed itself into family learning program with information literacy as a centerpiece. Founded in 1988 by some school teachers frustrated by the educational bureaucracy, the ALA has adopted a "social learning" model to improve educational outcomes. In 1996, ALA began to use the Internet as a literacy tool, and quickly discovered that it helped children with their reading, since the kids wanted to surf the Web and doing that made practicing reading part of a fun activity. From a "mom and pop" non-profit in 1988, the ALA has grown to a $1 million per year enterprise with over 50 computers used for education and Internet access.

In Washington, D.C., the SeeForever Foundation has an educational mission focused on at-risk youth, with a related goal to contribute to the revitalization of the Shaw neighborhood in the District. The SeeForever Foundation has established the Maya Angelou Public Charter School, serving approximately 80 students in a newly renovated building in Shaw. Many students live at the school and a central part of the curriculum is computer and Internet literacy. In addition to providing students essential course work, the school also offers practical exposure to real world employment. The school's "Student Tech Center" is one part of that; there students teach parents and siblings computer and Internet skills and engage in graphic design projects for neighborhood clients. Students are paid for teaching and receive evaluations of their work. The Maya Angelou students will also participate in a "Back Pack Technology Program" to reach out to neighborhood senior citizens. Supported by a $395,000 TOP grant, Maya Angelou students will provide training and support services to seniors in the Shaw area as students introduce them to the Internet. In the future, the SeeForever Foundation plans to aid neighborhood non-profits in obtaining Internet access and developing Web pages.

### c. Neighborhood Entrepreneurship

In Portland, a neighborhood community development corporation has used Internet and computer training as a tool to inject new life into the outer southeast part of the city. In 1995, the Neighborhood Pride Team (NPT) conducted a door-to-door survey to see what residents—mostly women, many raising children on their own—wanted from NPT. Two-thirds said that they wanted computer training and nearly half (43%) said they wanted to start a business in their home. Today, NPT has two full-time computer and Internet instructors and approximately 1,200 students pass through classes each year. A number of students have revived home based businesses in part by setting up Web pages and some people have discovered a knack for Web design that has become profitable undertakings for them. More widely, a group of artists known as the Trillium Artisans have put their wares online, thereby expanding the reach of their marketing efforts.

In Cleveland, a coalition of community activists called Digital Vision has spearheaded an effort to obtain from the city resources for city-wide computer "boot camps" for residents of low-income neighborhoods. The city has decided to channel $3 million of cable franchise fees to the boot camp program; the funds will be distributed to various community development corporations (CDCs) around the city. Part of the objective is to make neighborhood CDCs more vibrant centers of community activity, using computers and the Internet as a draw. At the same time, many activists believe that there is a spirit of entrepreneurship in many of these neighborhoods, and they hope that the Internet can encourage that in new ways. Whether people use the Web to start new businesses or enhance existing ones, Digital Vision activists recognize that technological literacy will improve people's prospects in the job market.

### d. Workforce Development

A number of the cities studied, such as Austin, Washington D.C., and Portland, are centers of high-tech economic activity. Most such regions have faced shortages of skilled workers in recent years. Activists in Austin and Washington have seized on this issue as a way to encourage access to information activity in low-income neighborhoods.

In Austin, city government—largely at the behest of community technology activists—launched a Telecommunity Partnership Initiative in 1998 to promote technology access through a $200,000 grant program. The city decided to devote the entire grant to a single organization focused on providing high-tech job training to Austin residents. An arm of the Austin Chamber of Commerce, the Capital Area Training Foundation, won the award and began providing training at one local high school in 1999. Since then the program has expanded to a second high school; unexpectedly high demand for training services led to the establishment of the second site. To meet other access needs in the community, the city started a second grants program in 2000, the Grants for Technology Opportunities Program (GTOPs). This $100,000 fund will make a series of smaller grants to community non-profits that may have specific focuses not relating to job training (i.e., technology access for women).

Byte Back is a Washington, D.C. non-profit that leverages a small paid staff, many volunteers, and a presence in numerous community based organizations throughout the city to offer two types of training to District residents. The first is basic computer and Internet literacy for anyone who walks into an organization where Byte Back courses are offered. Examples of such courses are introduction to Windows or basic Internet. This constitutes the bulk of Byte Back's foot traffic, with over 2,000 people having passed through one of these classes since Byte Back was founded. Students pay nominal fees for these courses, $10 for introductory ones and $25 for more advanced ones such as Power Point or HTML. As of the spring of 2001, Byte Back courses were available at 9 sites throughout the cities, ranging from Byte Back headquarters near Catholic University to a family center in a housing project, several churches, a Boys and Girls club, and a Catholic Charities facility. Classes are usually limited to about 10 people. For spring 2001, 560 students were participating in 70 Byte Back courses at sites throughout the city.

The second type of training is a year long program for what Byte Back calls their interns. This takes students to an advanced state of training that should enable them to gain employment in the information technology industry. Interns "test in" to the program based either on pre-existing computer aptitude or skills gained in basic Byte Back courses. Once accepted into the program, interns undertake a number of responsibilities. First, they commit to 30 hours a week to the program. Ten hours involve class time, ten hours are devoted to homework, and ten hours go to service at Byte Back headquarters or sites. Through their service to Byte Back, the interns essentially run the program. Byte Back has two full-time employees, but approximately 40 interns at any given time. With many sites throughout the city, a Byte Back intern serves as the technical support for a particular site's computers, in addition to teaching courses at the site. In addition to providing technical support, interns' responsibilities at the site also build technical skills for the job market. Interns also attend bi-monthly community meetings.

### e. Community Development

Needless to say, community technology projects have community development as a broad objective, and as we have seen, this plays out with different priorities in community technology projects, such as education or workforce development. However, some places have ambitious plans underway to use the Internet for community-building.

Nashville is using a TOP grant as part of a larger strategy to increase the voice of neighborhood groups in government affairs. Traditionally Nashville Metro government, which is a regional government combining Nashville and Davidson County, has not been noted for reaching out to neighborhood groups in the city. Nashville's current mayor, Bill Purcell, was elected in 1999 on a platform of reaching out to neighborhoods. Emboldened by a new commitment to neighborhoods, several community groups came together with Metro Planning Department to submit a grant proposal to TOP. The $477,000 award was made toward the end of 2000 for Nashville's "Designing a Community Online" project. The project seeks to address the following problems:

- Inaccessibility of public information—public information is often widely dispersed in Davidson County, and community groups often do not know where it is. There is no central office that might direct citizens to the right information.
- Community groups' lack of access to information makes it difficult for them to participate in shaping the community's future.
- Insufficient dissemination to the public of changes to zoning regulations and of development decisions.

To address these problems, the TOP project proposes a two-fold strategy of: a) assembling Metro content in user-friendly ways for the public, and b) increasing the number of public access sites throughout the Metro area. In terms of public information online, the project will put the following kinds of information online for the public: crime, land (e.g., floodplains, topography), historic properties and sale values, development plans, street plans, public transportation, population, and a variety of resources for the neighborhoods (boundaries, contact information, community-based social services). On the public access side, the grant proposes to purchase 75 computers, all connected to the Internet, and place them in 53 neighborhood and ethnic-based organizations in the Nashville area. This initiative is not a home-based or community Internet access project. Rather, the hope is that people in the neighborhood centers will serve essentially as Internet evangelists for individuals in their community.

Another effort explicitly oriented toward community building is the Austin Idea Network. This is a group of high-tech executives who want to use the wealth and talent of Austin's technology business community to reduce inequality citywide. Part of the Idea Network's program is to use information technology in schools; the plan is to provide computers and Internet access to an entire grade of students and teachers in Austin's public school system. The Idea Network may be a casualty of the dot-com shakeout; many of the network's founders are dot-com executives whose companies have gone out of business.

In other places, community development is seen as a long-term consequence of programs undertaken today. Portland's Neighborhood Pride Team has an "each one, teach one" approach to Internet and computer training. Out of the many people that come through training classes, NPT officials hope that a dozen or so will help others in their communities become computer literate. The Austin Free-Net's effort at long-term community building falls under the rubric of developing "community competence" with the Internet's help. This means increasing the community's capacity for helping itself through the knowledge and skills learned through the Internet. Such increased competence could take the form of better jobs for people in Austin or greater ease in finding places to live. Indeed, according to the TOP evaluation report of the Free-Net grant, increased self-esteem among community residents has been an important outcome of the grant. Austin Free Net's presence has grown to 34 sites around the city, up from 11 since its inception in 1995. The community-building story is much the same in Washington, D.C. Byte Back's job training and SeeForever's education missions are paramount, but both reach into the community. Byte Back does this by partnering with community organizations to provide services. SeeForever does this through its "back pack laptop" program to introduce the Internet to senior citizens.

# IV. THE INTERNET, CITIES, AND SOCIAL CAPITAL

The Internet is helping to change the "rules of the game" in various institutions within cities. In most cases, the Internet's effect is primarily catalytic. By prompting people to come together to plan how to use the Internet, the Internet's presence stimulates social networks and lays the groundwork for building new social capital. The development of Internet content as a result of the catalytic effect is a stronger signal that the Internet is building social capital. This has been less prevalent in the cities studied, but there are identifiable examples of it. In this concluding section, I summarize how the Internet's impact on social capital through the prism of the framework discussed at the outset, namely how the Internet affects social capital (catalyzing new social networks) and how social capital affects the Internet (through the creation of Internet content that serves the community). I also make recommendations on how to build upon early efforts that have had success in stimulating the creation of social capital.

### a. The Catalytic Effect of the Internet on Social Capital

In Cleveland, Bill Callahan, the director of the West Side CDC characterizes the Internet's impact on his organization in this way: "It wouldn't make sense to have this place without the Internet, and it wouldn't make sense to have the Internet in this neighborhood without this place."[8] In other words, this CDC needs to have Internet access in order to make it a place where people want to go, but merely providing Internet access to people in a disembodied fashion would not work. People need a place to learn about the Internet, and they also benefit from having an environment where they can find out who else in the neighborhood is on the Internet. In Callahan's CDC and others, the Internet has changed the pattern of foot traffic; people who probably otherwise would not visit a CDC are being drawn to it because of the Internet. With the Digital Vision coalition and the money from the city for computer boot camps throughout the city, Cleveland is in a position to expand the Internet's role in social networking.

In a number of other community organizations profiled, the catalytic effect of the Internet in changing foot traffic has drawn people to community organizations. The Neighborhood Pride Team revitalized itself by providing Internet and computer training to residents of a southeast Portland community. The Austin Learning Academy, due largely to the Internet, has become more than an after school program for kids, but rather a family learning center where the Internet is the centerpiece. And by partnering with existing community organizations in the District of Columbia, Byte Back's Internet training programs have added new dimensions to the missions of these non-profits. In these examples and others (e.g., the Austin Free-Net, D.C.'s See Forever project), the presence of the Internet increased foot traffic and served as a catalyst to social networking.

---

[8] Interview with Bill Callahan, director of West Side Community Development Corporation, Cleveland, Ohio, September 28, 2000.

### b. Content Development: Social Capital's Effect on the Internet

In several cities, there are indications that people have begun to begun to develop Internet content to help their communities or organizations operate more effectively. Some of this builds on existing stocks of social capital, while some is a direct outgrowth of Internet initiatives that first stimulated new networks of people. These people in turn have been instrumental in creating new content.

Portland's Neighborhood Pride Team again is a prominent example. NPT became a focal point for people interested in the Internet who subsequently created their own online content. The Trillium Artisans, who now market their wares on the Web, are an example of content creation, as are the people who have been through the NPT and then posted web pages to market home-based businesses. At the Austin Learning Academy, the creation of content is a central part of the "activity based" learning that is at the heart of ALA's teaching approach.

In Cleveland, theT2K.org portal for affordable housing providers is an example of Web content designed to improve service delivery for providers and clients of public housing. In the design phase, the Portland Area Housing Clearinghouse serves a similar purpose. Also in the design phase is Nashville's Designing a Community Online project; this project will provide access to the Internet for underserved populations and also create community Internet content. Further upstream is Austin's Grants for Technology Opportunities program that will fund Internet content development among non-profits. Although the development of Internet content by communities is not a widespread phenomenon, projects such as Portland's NPT indicate that it can have a substantial impact on people's lives when does occur.

### c. Social Capital and The City: Sustaining the "Internet Effect"

Even with signs that the Internet is having a positive impact on social capital, sustaining these impacts will be no easy task. Community technology programs typically struggle for resources and a key source for support from the public sector, the Technology Opportunities Program, is under constant budget pressure in Washington. Nonetheless, lessons from the five cities point to ways in which early efforts to capitalize on the Internet can be sustained.

**Bottom-up initiatives**: Almost invariably, the Internet projects in the five cities started because interested people in the community took the initiative. This underscores the "demand pull" nature of successful programs, rather than a "technology push" approach. In particular, the social networking approach employed in community technology programs have reached out into the community to determine community needs and respond to them. They do not impose community-computing programs on communities from the top-down. Portland's Neighborhood Pride Team is a good example of this, and there are similar examples in Austin, Cleveland, and Washington, D.C.

**Encourage catalysts**:[9] With the "Internet as catalyst" to building social capital a prominent theme in this report, a recommendation to encourage community catalysts is hardly a surprise. The bottom-up nature of most of the Internet initiatives has come about because individuals in the community have served as catalysts. Just because these people have taken the initiative does not mean that they and their initiatives do not need nurturing. Financial support is the most obvious, and probably most useful, form of encouragement, but publicity is another. The media could do a community service by focusing on how community groups are using the Internet for social purposes. This might help these programs obtain resources—financial, volunteer, and technical—that they otherwise could not.

**Public Funding**: Dollars from the coffers of local governments have played an important role in several cities. Cleveland and Austin have programs that channel public funds to community technology projects, although it is important to underline that the programs came about only after community technology activists had been running technology programs in the cities for some time. But as demand in the community for services expands, local government help is needed to meet it. Additionally, federal funding, in the form of U.S. Commerce Department's TOP grants, often is crucial to getting projects off the ground. There is still considerable demand for community computing programs and great need to wire local governments for better service delivery. With this high level of demand likely to be evident for some time into the future, cutting the TOP, as has been rumored, does not appear wise.

**Encourage "bridging" among groups**: In several cities, coalitions have been formed that have amounted to attempts to foster "bridging" social capital, that is, bringing advocates of low-income people into contact with the technology sector for community development. Cleveland's Digital Vision and Austin's Idea Network are two examples. Such initiatives hold significant promise, but the existence of them should not be seen as ends in themselves. These coalitions are partnerships among people and groups with differing outlooks and goals. Business leaders may see community-computing programs as way to increase the supply of skilled workers—something that they need quickly. Community activists may see the partnerships as part of long-term strategies to improve people's lives and foster civic engagement among forgotten members of the community. Recognizing these differences early is key to making bridging efforts work.

**Encourage experimentation**: Across the five cities, we have seen a number of different models for using the Internet for community purposes. In Washington, D.C., the non-profit Byte Back partners with existing social service agencies to provide Internet access. In Austin, the Free-Net, which receives some public support, also partners with community groups, but the city also funds job-training programs that focus on computer skills. There is no single solution to exploiting the Internet's potential and community leaders and policymakers should be aware of this. A willingness to tolerate multiple approaches should also be accompanied by a willingness to tolerate fits and starts in programs, and even failure. The lessons learned in the process can be as valuable as successful models that are often showcased as successes.

---

[9] This idea builds on a recommendation in the Kettering Foundation's report "Meaningful Chaos: How People Form Relationships with Public Concerns" available on the web at: http://fly.web.net/ccic/volsect/PE4A-meaningful_chaos.htm

The Internet brings new possibilities to cities and communities, from improved delivery of government services, to greater civic engagement, to new economic opportunities for regions. But cities also bring new possibilities to the Internet, as community leaders can bring content to the Internet that furthers a wide variety of community objectives. The reciprocal relationship between the Internet and places is how the "rules of the game" for institutions are shaped. There are early signs that the Internet can play a positive role in revitalizing city institutions. Patience, persistence, and resources will be needed over time to sustain and build upon these early successes.

# Appendix: Economic Profiles of the Five Cities

The five cities studied in this report have a variety of characteristics, with several being among the most highly wired cities in the United States (Austin, Portland, and Washington, DC), some being centers of high-tech manufacturing (Austin and Portland), others being service oriented (Nashville and Washington), and one traditional manufacturing center (Cleveland). The following tables present portraits of the five cities.

*Table A.1   Population and Internet indicators for case study cities.*

|  | Internet Penetration Rates at home or work (% of adults) (2000) | Internet Penetration Rates at home or work (% of adults) (1999) | Population (2000) | Population Growth in percentage ('90-'00) | Median Age in years (1999) |
|---|---|---|---|---|---|
| **Washington, DC** | 73 | 71 | 4,815,581 | 14.1 | 33.5 |
| **Austin** | 69 | 64 | 1,186,279 | 40.2 | 30.2 |
| **Portland** | 61 | 57 | 1,870,730 | 23.5 | 34.7 |
| **Nashville** | 50 | 50 | 1,187,521 | 20.6 | 33.5 |
| **Cleveland** | 48 | 42 | 2,217,174 | 0.7 | 33.5 |

*Table A.2   Economic Structure of Cities (% Employment in Each Sector, excluding Public Administration)*

| SECTOR | Year | Con-struction | Manu-facturing | Distrib-utive Services | Sales (Whole-sale + Retail) | Producer Services | Health & Educa-tion | Other Services |
|---|---|---|---|---|---|---|---|---|
| **Cuyahoga County** | 1987 | 3.9 | 25.2* | 5.4 | 27.2 | 15.4 | 12.2 | 10.7 |
| **(Cleveland-Lorain-Elyria, OH)** | 1997 | 3.4 | 19.3* | 4.8 | 25.8 | 22.1 | 14.9 | 9.7 |
| **Davidson County** | 1987 | 7.1 | 16.6 | 6.8 | 29.2 | 15.7 | 12.0 | 12.6 |
| **(Nashville, TN)** | 1997 | 5.4 | 10.9 | 7.2 | 28.1 | 19.2 | 16.1 | 13.0 |
| **District of Columbia** | 1987 | 2.8 | 4.3 | 5.5 | 16.7 | 29.2 | 17.7 | 23.8 |
| **(Washington, DC)** | 1997 | 1.6 | 3.1 | 5.0 | 13.6 | 31.8 | 21.6 | 23.2 |
| **Multnomah County** | 1987 | 3.7 | 15.5 | 10.1 | 29.3 | 19.1 | 10.6 | 11.8 |
| **(Portland-Vancouver, OR)** | 1997 | 5.2 | 13.2 | 8.9 | 27.1 | 22.1 | 12.7 | 10.8 |
| **Travis County** | 1987 | 5.6 | 15.4** | 4.3 | 30.4 | 22.0 | 7.7 | 14.6 |
| **(Austin-San Marcos, TX)** | 1997 | 5.5 | 16.2** | 4.2 | 26.6 | 26.9 | 10.1 | 10.6 |
| **United States** | 1987 | 5.8 | 22.7 | 6.1 | 28.9 | 15.0 | 10.2 | 11.3 |
|  | 1997 | 5.3 | 18.0 | 6.0 | 27.9 | 18.9 | 13.1 | 10.8 |

\* Percentage of employment in machinery was approximately 13.0 % in 1987 and 10.4 % in 1997.

\*\* Percentage of employment in electronic and other electric equipment was approximately 3.9 % in 1987 and 7.7 % in 1997.

**Data Source: County Business Patterns, "http://fisher.lib.virginia.edu/cbp/"**